\documentclass[aps,prl,floats,amssymb,twocolumn,letterpaper,groupedaddress,
showkeys,floatfix,twoside]{revtex4}
\usepackage{graphicx}
\usepackage{array}
\usepackage{color}
\usepackage{epsfig}
\usepackage{float}

\begin{document}

\newcommand{\pe}{\psi}
\def\d{\delta} 
\def\ds{\displaystyle} 
\def\e{{\epsilon}} 
\def\eb{\bar{\eta}}  
\def\enorm#1{\|#1\|_2} 
\def\Fp{F^\prime}  
\def\fishpack{{FISHPACK}} 
\def\fortran{{FORTRAN}} 
\def\gmres{{GMRES}} 
\def\gmresm{{\rm GMRES($m$)}} 
\def\Kc{{\cal K}} 
\def\norm#1{\|#1\|} 
\def\wb{{\bar w}} 
\def\zb{{\bar z}} 

% some definitions of bold math italics to make typing easier.
% They are used in the corollary.

\def\bfE{\mbox{\boldmath$E$}}
\def\bfG{\mbox{\boldmath$G$}}
\definecolor{DarkBlue}{rgb}{0,0,0.5}
\definecolor{DarkRed}{rgb}{0.5,0,0}

\renewcommand{\topfraction}{0.95}
\renewcommand{\dbltopfraction}{0.95}
\renewcommand{\textfraction}{0.05}

\title{
Analytic Formulas for the Orientation Dependence of Step Stiffness and Line Tension:
Key Ingredients for Numerical Modeling
}

\author{T. J. Stasevich}
\email[]{tim\_stasevich@hotmail.com}
\author{T. L. Einstein}
\email[]{einstein@umd.edu}
\homepage[]{http://www2.physics.umd.edu/~einstein/}
\affiliation{Department of Physics, University of Maryland, College Park, MD 20742-4111
%\\ $^\star$Now at: Fluorescence Imaging Facility, National Cancer Institute, 
%National Institutes of Health, 9000 Rockville Pike, 
%Bethesda, MD 20892
} 

\begin{abstract}
We present explicit analytic, twice-differentiable expressions for the temperature-dependent anisotropic step line tension and step stiffness for the two principal surfaces of face-centered-cubic crystals, the square \{001\} and the hexagonal \{111\}.  These expressions improve on simple expressions that are valid only for low temperatures and away from singular orientations.  They are well suited for implementation into numerical methods such as finite-element simulation of step evolution.
\\{}\\
AMS codes: 82B24, 
80M10, %Classical thermo: Finite element methods 
35Q99, %Equations of mathematical physics and other areas of application: None of the above, but in this section
82C05, %Classical dynamic and nonequilibrium statistical mechanics (general)
76M28
\end{abstract}

\keywords{ 
step stiffness, step line tension, anisotropy, numerical modeling, finite-element simulation, step dynamics
}

\date{November 27, 2006}

\maketitle

\section{Introduction}

Study of stepped crystalline surfaces offers an exquisite combination of questions of fundamental physical interest and importance for technological development. 
These surfaces play a central role in the modern electronics industry, where they serve  
as templates in the computer chip manufacturing process.
The ability to model and quantify the evolution of such surfaces and ultimately engineer surface structures over a wide range of length and time scales will be essential for designing the next generation of computer chip components \cite{barth}.
Even on the flattest of surfaces, e.g.\ Si(111), steps are inevitable on the submicron scale due to minor misorientations from facet directions.  For metallic surfaces, steps appear on the scale of 100nm.  In some cases they provide a mechanism of stress relief \cite{noz}.  

In  applications to electronics, template 
surfaces are often grown epitaxially: material, usually in the form of 
atoms or molecules, is sputtered onto a crystalline substrate; the goal is to grow layer by layer with as few defects as possible.  Many excellent books discuss this general problem from a variety of perspectives (e.g.,  \cite{AP-JV,TM-JK,modl}).
 If one starts with a flat surface, growth begins at random nucleation sites.  Islands expand around these sites.  Eventually the boundaries of the islands meet.  If the atoms in two abutting islands are in different domains, then a domain wall forms between them.  Such walls persist during growth and cannot readily be annealed away.  Instead, it is advisable to begin with a substrate that is intentionally slightly misoriented (typically by a few degrees) from flat, high-symmetry orientation.  Such surfaces are called ``vicinal" since they are in the vicinity of the facet orientation.  
They consist of a series of atomically flat terraces, separated from one another by 
surface steps---boundaries where the surface height changes by an atomic unit. 
Ordinarily the adsorption energy of a deposited atom is greatest at the crease at the lower edge of a step, since there it can bind to the largest number of other atoms.  Thus, if the temperature is high enough so that the atoms diffuse relatively rapidly and the flux low enough, the atoms will be more likely to attach to the step edge than to meet another deposited atom to form a nucleation center on a terrace.  Then, in this regime of ``step-flow growth," the steps will gradually move across the terrace until, after deposition of a monolayer, the surface looks very similar to the initial vicinal surface, only with one more layer.   

Since steps play such a fundamental role, it is crucial to understand their properties and especially to clarify the few basic parameters that determine their behavior.  The lowest-energy excitation of a stepped surface is the kink, a unit deviation perpendicular to the mean direction of the step.  Since the energy to create an isolated atom or vacancy defect on a terrace is several 
times that to create a kink, kinks are the 
predominant defect on equilibrated surfaces at low temperatures.  While these kinks cost energy, they contribute to the entropy in the usual way, so that in equilibrium this competition leads to the step free energy per length $\beta$, or line tension \cite{ibachstress}.  In an expansion of the projected free energy per area of a vicinal surface (i.e., the surface free energy per area of the vicinal surface projected onto the terrace plane), $\beta$ is the coefficient of the density of steps.  (At the roughening transition of the terrace plane, $\beta$ vanishes and the projected free energy per area becomes quadratic in the density of steps; the vicinal surface is in this technical sense rough already at the lower temperatures under consideration.) 
In other words, the line tension indicates the extra energy associated with a unit length of a step.

Steps need not run along high-symmetry directions, i.e. the surface normal (or azimuthal misorientation) can be at an  
arbitrary polar angle $\theta$.  Furthermore, the border of a single-layer island (or vacancy island) is just a step that is a closed curve---somewhere between circular and polygonal---rather than a nearly straight line.  
The celebrated Wulff construction \cite{WortisVII} uses $\beta(\theta)$ to determine the equilibrium crystal shape that minimizes the free energy at constant area.

The thermal excitation of kinks along the step leads to meandering.  Such fluctuations are constrained by the stiffness $\tilde\beta(\theta) \equiv \beta(\theta) + \beta^{\prime\prime}(\theta)$, which weights the squared slope of the step relative to its mean direction \cite{FFW,stiff,TLE-ibach}.  (There is an analogy between the ensemble of spatial configurations of the steps in 2D (two dimensions) and the world lines of particles evolving in 1D; in this picture, the deviation of the slope is analogous to velocity, so the wandering corresponds to kinetic energy, with stiffness playing the role of mass.)  Furthermore, it is the stiffness that weights the curvature in the Gibbs-Thomson term, the curvature contribution to the step or interface energy.  Thus, in many ways $\tilde\beta(\theta)$ is more fundamental than $\beta(\theta)$ and in some situations is better defined \cite{akutsustiff}.  

Accordingly, 
stiffness is one of the three parameters of the step continuum model, which retains distinct steps but coarse-grains them into continuous ``strings."  The other two parameters are the strength of the inverse-separation-squared repulsion between steps and the characteristic rate of the mechanism dominating step kinetics.  Then, at the mesoscopic scale, such surfaces can be envisioned as a collection of 
steps separating  
atomically flat terraces and effectively tracing out surface contour plots.  By tracking the net movement, fluctuations, and interactions of steps, the evolution of the 
entire surface can be monitored.  The actual motion of atoms---at much smaller length and time scales---which underlies this behavior appears only in the parameters, which can be estimated from either calculation (or measurement) of atomic processes or by fitting to data at the mesoscale.  This idea, at the heart of the step-continuum model \cite{jeongwms}, 
turns out to be an extremely efficient way of tracking   
surface evolution over a wide range of experimentally relevant length and time scales. 
This picture has been used to successfully account for a broad range of step properties, such as mound decay, step and island fluctuations, cluster diffusion, electromigration, and ripening \cite{jeongwms}.  Since one can never be certain of including all significant atomistic-scale processes, the most meaningful test of the step continuum model is the check that the same set of parameters describes quantitatively all these varied phenomena.  One can then return to the microscopic picture, invoke a simple model, and fit the key model parameters as {\it effective} values to reproduce the mesoscopic behavior.  If reliable energetic calculations give agreement with these predictions, all the better.

In this paper we concentrate on the two densest, highest-symmetry faces of an fcc crystal, namely \{100\} and 
\{111\}, which have square and hexagonal symmetry, respectively.  (For \{111\} surfaces, if one imagines adsorption into both kinds of 3-fold sites [fcc and hcp], one has a honeycomb.  This feature is unimportant for what we consider here.)  Late transition and noble metals have fcc crystal structure and make good substrates for the sorts of experiments envisioned here, since they are relatively soft, with atomic motion occurring adequately for equilibration at room temperature or at somewhat higher.  As the close-packed Bravais structure, there is less angular dependence on bonding, making near-neighbor models better approximations than for bcc metals.  Note that \{100\} and \{111\} faces of all cubic Bravais crystals have square and hexagonal symmetry, respectively.  So does Si and other systems with the diamond structure.

If one assumes that step adatoms interact with only nearest-neighbors (NN) 
or next-nearest-neighbors (NNN), then it is possible to
derive exact solutions for the line tension based on the Ising or solid-on-solid 
(SOS) models.  
These solutions are implicit, however, making their 
implementation into numerical simulations time-consuming and computationally demanding, 
particularly when dealing with the stiffness, which requires
two additional derivatives of the implicit line tension.  
For simplicity, then, numerical studies often \cite{iso1,iso2} (though by no means always \cite{weeks}) assume an isotropic line tension and stiffness.  Except at high temperatures where an island structure is nearly circular, this approximation turns out to be poor, especially near facet orientations, where the line tension is notably smaller, and would be much smaller if there is a quasi-straight-edge (2D ``facet"---at macroscopic scales the island is rounded at finite temperature \cite{rottman3d}).  
For the stiffness, the problem is more severe, since $\beta^{\prime\prime}(\theta)$ is large near these special directions (though not infinite as it would be in 3D), leading to a stiffness much greater than the value at general orientations.  Since it is easier to compute $\tilde\beta(\theta)$ in such high-symmetry directions, such values have been used to characterize $\tilde\beta(\theta)$ at general $\theta$, thereby considerably overestimating the typical stiffness.

The next simplest approximation assumes a sinusoidal variation reflecting the substrate symmetry \cite{cos}.  Again, there are shortcomings to this procedure, especially near 
 facet orientations, about which polar plots of the line tension as a function of angle 
reveal sharp, cusp-like minima, implying the stiffness blows up (since the curvature of a cusp
 is infinite).
Such temperature-independent simplifications preclude 
quantitative comparisons with experiment \cite{dielu}.

In this paper we construct expressions for $\beta(\theta)$ and $\tilde{\beta}(\theta)$ that are well behaved analytically, being continuous and twice differentiable and that give an accurate accounting at all orientations and relevant temperatures.  For convenience, all derived formulas are summarized in Table~5.1.  
While not especially simple, they are straightforward to construct and easy to implement in numerical codes such as used in finite-element investigations \cite{voigt1,voigt2}, making quantitative comparisons with dynamic experiments 
possible.  We thus expect our results to be widely applicable.

Our approach begins with simple, low-temperature formulas for 
the orientation dependence, on face-centered-cubic (fcc) surfaces,  
of the \{001\} and \{111\} stiffness  and line tension   
that we derived in two recent papers \cite{tjs1,tjs2}.  (This approach is rooted in the lattice-gas perspective, so is complementary to Shenoy and Ciobanu's study of stiffness anisotropy based on elasticity theory \cite{ShenoyAni}.) 
Our formulas assume the step fluctuations are dominated by the rearrangement of
 geometrically forced kinks---kinks that are 
not thermally activated.  At temperatures low compared to the surface roughening temperature
 (for noble metal surfaces, such as Ag and Cu, room temperature is considered ``low''),  
the formulas only fail for steps having 
a negligible number of forced kinks; that is, steps oriented very close to the high-symmetry direction.  When the step angle is 
exactly 0$^\circ$ (aligned with the high-symmetry direction), 
the formulas predict a cusp in the line-tension and an
infinite step stiffness.  Here we correct for the non-analytic behavior by
splicing our simple, low-temperature formulas with small-angle expansions 
of the exact, implicit solutions based on the Ising and SOS models.

In the following section, we describe the details of a general expansion for the  
stiffness and line tension that is continuous and twice-differentiable.  
In sections III and IV, we apply this expansion 
to fcc \{111\} and \{001\} surfaces, respectively, to derive surface-specific 
formulas for the stiffness and line tension.  
In the final section, we offer concluding 
remarks as well as a synopsis of the derived expressions.

\section{Explicit Analytic Approximation}
\begin{figure*}[ht]
\begin{center}
\includegraphics[width=10.2cm]{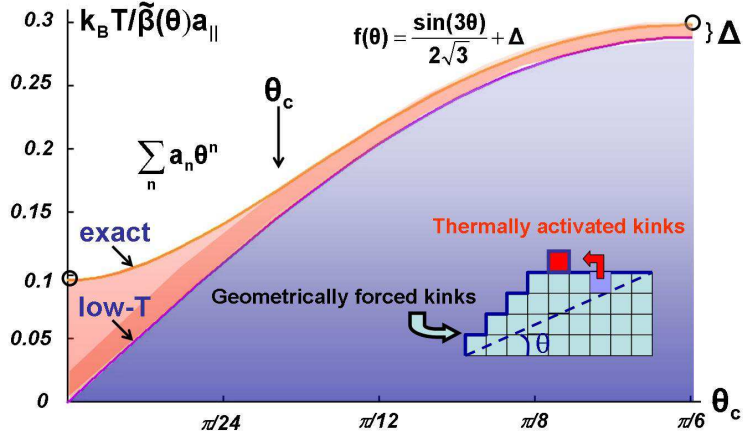}
\end{center}
\caption{(Color online) 
The contributions to the step stiffness can be decomposed into parts originating from  
geometrically forced kinks (lower blue region bounded from above by the line labeled 
``low-T'') and thermally activated kinks (the remaining red region, bounded from above by the 
line labeled ``exact'').  
At relatively low-temperatures, the \{111\} step stiffness is well approximated 
at angles greater than $\theta_c$ by a relatively simple, explicit function 
$f(\theta)$, since the thermal part is evidently insensitive to angle.  To account for all angles, the formula can be spliced 
with a small-angle expansion of the exact NN Ising model solution (from which 
explicit forms 
for the stiffness can be obtained at  $\theta=0$ and at $\pi/6$, depicted here
 by hollow circles).  The solution at $\pi/6$ is used to determine $\Delta$. The expansion coefficients $a_n$ are obtained by matching the solutions at $\theta=0$ and $\theta_c$.  The inset depicts a step edge from above.  Each square represents an adatom which is part of 
the step edge.  The upper-most square represents a thermally excited adatom, which 
forms four thermally-activated kinks.  The remaining kinks are geometrically forced---they 
must be present to give the step edge an overall angle $\theta$. 
}
{\label{fig:smallAngleExpSample}}
\end{figure*}

At the microscopic level, the step stiffness and line tension 
arise from the energy and rearrangement of step edge 
kinks.  
It is therefore natural to decompose $\tilde{\beta}(\theta)$ and 
$\beta(\theta)$ into two contributions: one part originating from geometrically 
forced kinks and one part from thermally activated kinks.    
Geometrically forced kinks, depicted in the inset of  
Fig.~\ref{fig:smallAngleExpSample}, are present at all temperatures, and  
give the step an overall orientation $\theta$.  The further $\theta$ is from the 
high-symmetry direction, the greater the number of 
geometrically forced kinks.
Thus, at lower temperatures, 
as long as the orientation angle of a step is    
{\it greater} than some small, temperature-dependent cross-over angle  
$\theta_c$, there are many geometrically forced kinks and 
 relatively few thermally activated kinks, suggesting   
 $\tilde{\beta}(\theta)$ and $\beta(\theta)$ can be well described by formulas 
based on geometrically forced kinks alone.

As an example, we have recently derived \cite{tjs1} 
a remarkably simple, low-temperature formula for 
the \{111\} step stiffness assuming only NN adatom interactions and geometrically 
forced kinks:
\begin{equation}
  \label{eq:111stiffLowT}
  \frac{k_B T}{\tilde{\beta}(\theta)}\approx \frac{\sin(3 \theta)}{2\sqrt{3}}.
\end{equation}

%%%%%%%%%%%%%%%%%%%%%%%%%%%%%%%%%%%%%%%%%%%%%%%%%%%%%%%%%%%%%%%%%%%%%%%%%%%%%%%%%

%%%%%%%%%%%%%%%%%%%%%%%%%%%%%%%%%%%%%%%%%%%%%%%%%%%%%%%%%%%%%%%%%%%%%%%%%%%%%%%%%
At sufficiently low (but experimentally relevant) temperatures, 
the formula works well for steps at nearly all angles, but 
predicts an infinite stiffness when $\theta=0$.  
Fortunately, the exact, implicit solution 
based on the NN Ising model can be explicitly written for steps having this orientation.  
We can therefore expand the exact solution about $\theta=0$ and splice it with 
our low-temperature solution at $\theta_c$, thereby 
producing an explicit form for $\tilde{\beta}(\theta)$ 
valid at all angles.  This idea is illustrated in Fig.~(\ref{fig:smallAngleExpSample}).
Here, an additional orientation-independent contribution to the stiffness from 
thermally activated kinks $\Delta$ is also included for completeness.  Similar to 
high-symmetry steps, the stiffness of 
maximally kinked steps ($\theta=\pi/6$) can be exactly obtained from the 
NN Ising model, so that $\Delta$ can  
be determined explicitly.

To generalize this approach, we assume $\tilde{\beta}(\theta)$ and $\beta(\theta)$ are 
well described at angles {\it greater} than $\theta_c$ by simple, analytic functions 
representing contributions from geometrically forced kinks.  Explicit forms for these 
functions \cite{tjs1,tjs2} 
will be discussed later.  For now, to be general, we 
simply write them as $f(\theta)$.  
  
At sufficiently low
temperatures, $\theta_c$ is small, so we may accurately 
represent $\beta(\theta)$ and the inverse stiffness 
$\tilde{\beta}^{-1}(\theta)$ at angles {\it less} than $\theta_c$ 
using small-angle expansions. (We expand the inverse   
stiffness because, in the $\theta$=0 limit, it vanishes at low temperatures, making it mathematically better behaved than the stiffness itself, which diverges).  Specifically, 
we construct an approximant $X(\theta)$ to represent the dimensionless form of the function we 
wish to expand---either $\beta(\theta) a_{||}/(k_B T)$ 
or $k_B T/(\tilde{\beta}(\theta) a_{||})$, 
where $a_{||}$ is the close-packed distance between atoms (i.e. the atomic diameter), and $k_B T$ is the Boltzmann 
energy---we define
\begin{eqnarray}
  \label{eq:expand1}
    X(\theta) &:=& \left\{ 
      \begin{array}{ll}
         \sum_{n=0}^{2 N-1} a_n~ \theta^n & \mbox{if $\theta < \theta_c$} \\
         f(\theta) & \mbox{if $\theta \geq \theta_c$} 
       \end{array}. \right.
 \end{eqnarray}
To fully specify this function, we must find the appropriate expansion coefficients, $a_n$.  
We obtain their values by matching Eq.~(\ref{eq:expand1}) and 
its higher order 
derivatives with the exact solutions at $\theta$=0 (which can be systematically obtained)
 and the approximate (yet accurate) solutions obtained from $f(\theta)$
 at $\theta=\theta_c$, analogous to performing a spline fit \cite{spline}.  Specifically, for the boundary conditions at $\theta=0$, we have
\begin{eqnarray}
  \label{eq:bca}
    a_n &=& 
       \frac{\partial^n_\theta X(0)}{n!},~~~~~ n<N 
 \end{eqnarray}
where $\partial^n_\theta X(0) \equiv \partial^n X(\theta)/\partial \theta^n|_{\theta=0}$.  
The remaining $N$ coefficients are found from the boundary conditions 
at $\theta=\theta_c$, which form a set of $N$ coupled linear equations:
\begin{eqnarray}
  \label{eq:bc2a}
    \sum_{n=N}^{2N-1} \frac{n!}{(n-m)!} a_n \theta_c^{n-m} &=& 
    \partial^m_\theta f(\theta_c) ,~~~ %m<N, 
\end{eqnarray}
where $m= 0,1,2,...,N-1.$

For use in continuum models, 
$\tilde{\beta}(\theta)$ should be continuous and 
twice-differentiable.  To ensure the second derivative remains continuous at 
$\theta=\theta_c$, this requires, at minimum, $N=3$.  In this case, 
Eqs.~(\ref{eq:bc2a}) are simultaneously solved to give:
\begin{eqnarray}
  \label{eq:bc}
    a_3&=&\frac{20 (f-X)-8 f' ~\theta_c + (f''-3 X'') ~\theta_c^2}{2 ~\theta_c^3}\\
    a_4&=&\frac{-30 (f-X)+14 f' ~\theta_c - (2 f''-3 X'') ~\theta_c^2}{2 ~\theta_c^4}\\
\label{eq:bcfinal}
    a_5&=&\frac{12 (f-X) -6 f' ~\theta_c + (f''- X'') ~\theta_c^2}{2 ~\theta_c^5},
\end{eqnarray}
where the prime represents differentiation with respect to $\theta$; for brevity we write $f \equiv f(\theta_c)$ and 
$X \equiv X(0)$.  Note we have also used Eq.~(\ref{eq:bca}), which 
implies $a_0 = X$, $a_1 = X'$, and $a_2 = X''/2$.  Because both the line tension 
and the stiffness are continuous and symmetric about $\theta$=0, we know that $a_1=X'=0$.   
In the remaining sections we apply this approximation to specific cases where explicit forms for 
$X$ and $f$ can be obtained.

\section{\{111\} Surfaces with NN Interactions}
For \{111\} surfaces with only NN adatom interactions, Zia found an implicit form for the full orientation dependence of the 
step line tension \cite{zia}: 
\begin{equation}
  \label{eq:ziaSol}
  \frac{\beta a_{||}}{k_B T}=\eta_0(\theta)\psi_1(\theta,T/T_c)+
               \eta_-(\theta)\psi_2(\theta,T/T_c),\\
\end{equation}
where $\eta_0(\theta)\equiv (2/\sqrt{3})~\sin(\theta)$, $\eta_{\pm}(\theta) \equiv \cos(\theta)
\pm (1/\sqrt{3})~\sin(\theta)$. Here $T_c$ is the critical temperature of the NN lattice-gas model. The $\psi$'s are solutions of the pair of simultaneous equations 
for the angular constraint,
\begin{equation}
  \label{eq:angConst}
  \frac{\sinh(\psi_1-\frac{1}{2}\psi_2)\cosh(\frac{1}{2}\psi_2)}{\sinh(\psi_2-\frac{1}{2}\psi_1)\cosh(\frac{1}{2}\psi_1)}=\frac{\eta_0}{\eta_-}, \\  
 \end{equation}
and the thermal constraint,
\begin{equation}
  \label{eq:thermConst}
\cosh \psi_1+\cosh \psi_2 + \cosh (\psi_1-\psi_2)=\frac{y^2-3}{2},
\end{equation}
where $y \equiv \sqrt{(3z+1)/z(1-z)}$ and $z\equiv3^{-T_c/T}$.  
The latter can be rewritten $z \equiv \exp(-2\epsilon_k/k_BT)$, where $\epsilon_k$ is the energy of a kink on a close-packed step and 

%%%%%%%%%%%%%%%%%%%%%%%%%%%%%%%%%%%%%%%%%%%%%%%%%%%%%%%%%%%%%%%%%%%%%%%%%%%%%%%%%
\begin{figure}[ht]
\begin{center}
\includegraphics[width=6.4cm]{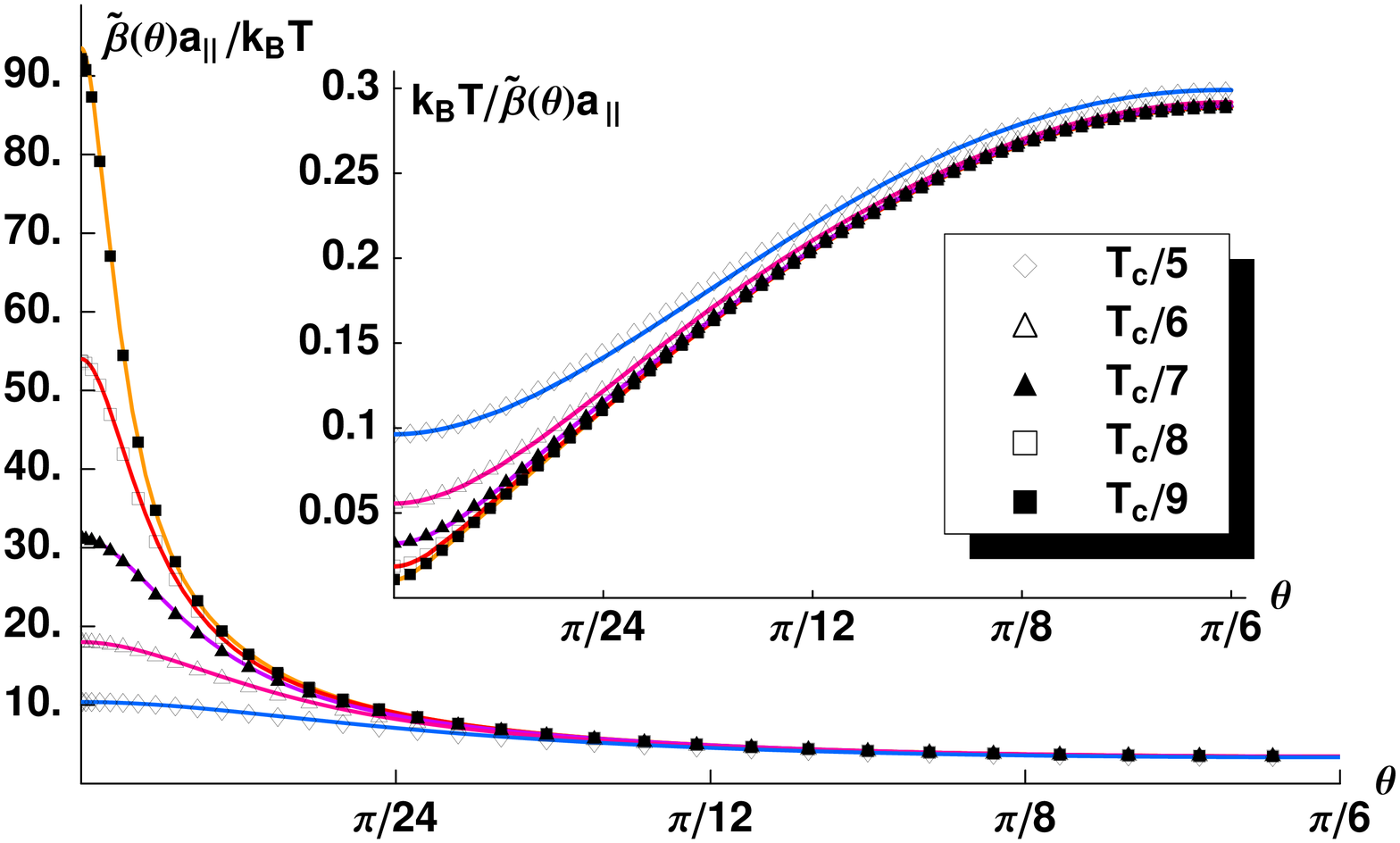}
\includegraphics[width=6.4cm]{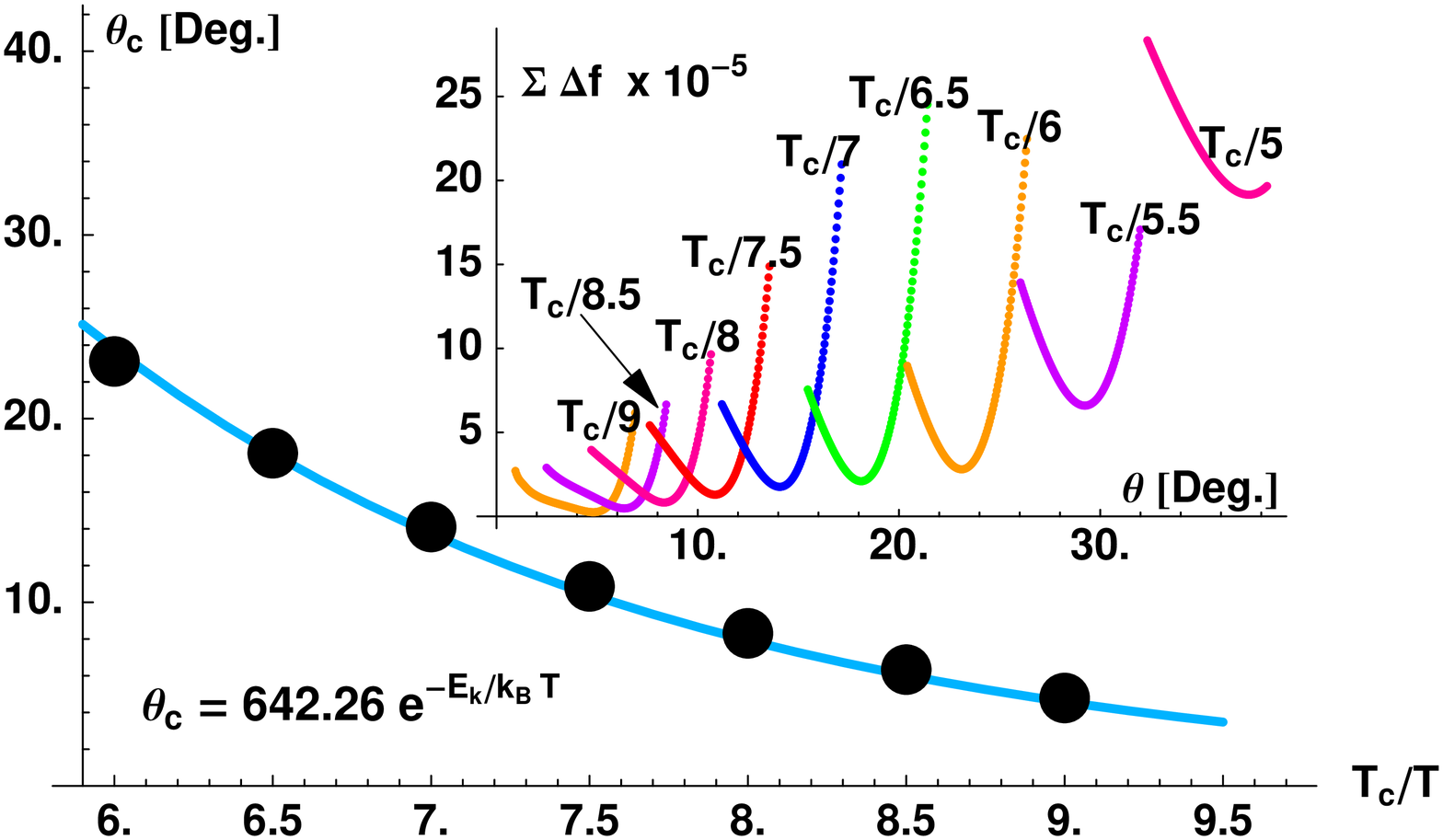}
\end{center}
\caption{(Color online) In the left plot, the 
orientation dependence of our explicit approximation  
for the \{111\} step stiffness (solid lines) and its inverse (inset, solid lines) are  
compared to the exact, implicit solutions (symbols).  Because of the six-fold 
symmetry of the solution, only the positive half of the first sextant is shown 
(the negative half is mirror-symmetric).  The right plot shows
the values used for $\theta_c$ (solid dots) in the construction of the left figure 
and the  corresponding exponential fit (solid line), good over the temperature range 
of interest.  The fit is expressed in terms of the kink energy $\epsilon_k$, which 
is related to $T_c$ by Eq.~(\ref{eq:kinke111}).
The inset shows the sum of squared vertical deviations ($\sum \hat{\chi}^2$) 
versus angle in a least square fit 
for $\theta_c$.
At each temperature, $\theta_c$ is the angle that 
minimizes this sum.  
}
{\label{fig:allAngSti}}
\end{figure}
%%%%%%%%%%%%%%%%%%%%%%%%%%%%%%%%%%%%%%%%%%%%%%%%%%%%%%%%%%%%%%%%%%%%%%%%%%%%%%%%%

\begin{equation}
  \label{eq:kinke111}
\frac{\epsilon_k}{k_B T_c} = \ln{\sqrt{3}}    
\end{equation}

From Eqs.~(\ref{eq:angConst},\ref{eq:thermConst}) it follows that
\begin{equation}
  \label{eq:psiZero}
  \psi_1(0)=\frac{1}{2}\psi_2(0)=\cosh^{-1}\left(\frac{y-1}{2}\right).
\end{equation}
With $\psi_1(0)$ and $\psi_2(0)$ in hand, we can differentiate the constraints,
Eqs.~(\ref{eq:angConst},\ref{eq:thermConst}), set $\theta=0$, and systematically solve 
for all the 
higher order derivatives of the $\psi$'s, which, according to Eq.~(\ref{eq:ziaSol}), 
are sufficient to find the higher order derivatives of $\beta$.  We will utilize these 
higher order derivatives to derive explicit, analytic approximations for the stiffness and 
line tension.  

\subsection{Step Stiffness}

In this case, $X(\theta) \equiv k_B T/(\tilde{\beta}(\theta) a_{||})$, 
which is six-fold symmetric for \{111\} surfaces with only NN adatom interactions.  
To utilize our explicit analytic approximation, we require 
$f(\theta)$---the contribution to the reduced 
stiffness from geometrically forced kinks---which, 
in the first sextant ($-\pi/6$ to $\pi/6$), takes a relatively simple form \cite{tjs2}:
\begin{equation}
  \label{eq:fstiff}
  f(\theta) = \frac{1}{2\sqrt{3}} \left( \sin (3 \theta) + \frac{3+y^2}{\sqrt{y^4-10 y^2+9}}-1\right).
\end{equation}
The last two terms, called $\Delta$ in Fig.~\ref{fig:smallAngleExpSample}, are included to ensure $f(\theta)$ matches the exact solution 
for steps with orientation angle $\theta=\pi/6$.  The physical origin of the $\Delta$ terms is the thermal fluctuations of a maximally kinked step.  Such fluctuations are relatively inexpensive in terms of energy.  They dominate the fluctuation contribution while a significant fraction of the step is not close-packed, so that the thermal contribution for such orientations is relatively independent of orientation.  Since only the first term 
has any $\theta$ dependence, $f'$ and $f''$ are simple to calculate.    

Now only $X$ and its first two derivatives need to be determined.  As mentioned 
in the preceding section, these can be systematically determined.  In particular, 
we find (see Eq.~(23) for a derivation of X in our earlier paper \cite{tjs2}):
\begin{eqnarray}
  X &\equiv& \frac{k_B T}{a_{||} \tilde{\beta}(0)} = \frac{3(y-1)}{2 y \sqrt{y^2-2y-3}}, \label{eq:xtri}\\
   X'&=&0, \\
   X'' &=& \frac{y^3-2y^2-15y+36}{2(y-1)\sqrt{y^2-2y-3}}. 
\end{eqnarray}
Of course, based on symmetry, we already knew that $X'=0$.  

By combining the functional forms for $f$ and $X$ and their derivatives with 
Eqs.~(\ref{eq:expand1}-\ref{eq:bcfinal}), we can plot the stiffness and compare it 
to the numerically evaluated exact solution.  
We show this comparison in Fig.~\ref{fig:allAngSti}, where 
$\theta_c$ was determined at a variety of temperatures 
by doing least square fits to the exact solution.  
The agreement shown in Fig.~\ref{fig:allAngSti} is very good at low-temperatures and 
is quite reasonable at temperatures all the way 
up to $T_c/5$.  (This behavior is remarkable since slightly above $T_c/5.5$, $\theta_c$ becomes greater than $30^\circ$, i.e., the power series is used for the entire range of orientations.  Once  $|\theta_c| > 30^\circ$, the slope of $k_B T/(\tilde{\beta}(\theta) a_{||})$ no longer vanishes at $30^\circ$.) 
At higher temperatures, the angular dependence becomes negligible, so $\tilde\beta(\theta)$ become 
isotropic. 

The right plot in Fig.~\ref{fig:allAngSti} 
shows the values used for $\theta_c$, along with an exponential fit:

\begin{equation}
  \label{eq:critang111}
  \theta_c(T)\approx 11.2 \, \exp(-\epsilon_k/k_B T) = 642[{}^\circ] (\sqrt{3})^{-T_c/T}.
\end{equation}
\noindent The second form uses the units in Fig.~\ref{fig:allAngSti}, 
reexpressing the prefactor in degrees and the exponent in $T_c/T$.  The Arrhenius decay  reflects the importance of thermally-activated kinks for $|\theta| < \theta_c$.
%  It is just the angle below which thermally activated kinks become important.  

\subsection{Step Line Tension}
We follow the same procedure for the line tension.  In this case $X(\theta) \equiv \beta(\theta)a_{||}/k_B T$.  Corresponding to the $T\! =\! 0$ divergence of the stiffness at $\theta \! =\! 0$ is a cusp in the line tension, indicating a facet in the equilibrium shape.  At finite $T$ the cusp, like the divergence, vanishing since a facet on a 2D structure corresponds to 1D long-range order. 
Specifically, the contribution (in the first sextant) to the line tension from geometrically forced kinks is 
fairly simple \cite{tjs1}:
\begin{equation}
  \label{eq:fkinkslt}
  f(\theta) = -\eta_{+} \ln z -\eta_+ \ln \eta_+ + 
                  \eta_- \ln \eta_- + \eta_0 \ln \eta_0 . 
\end{equation}
%%%%%%%%%%%%%%%%%%%%%%%%%%%%%%%%%%%%%%%%%%%%%%%%%%%%%%%%%%%%%%%%%%%%%%%%%%%%%%%%%
\begin{figure}[h]
\begin{center}
\includegraphics[width=8.2cm]{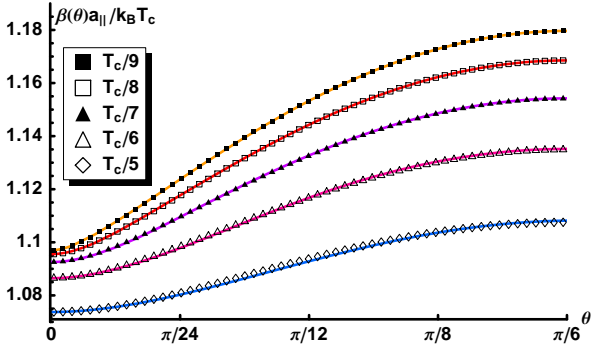}
\end{center}
\caption{(Color online) The orientation dependence of the explicit approximation 
for the  \{111\} line tension (solid lines) is compared with the numerically evaluated 
exact result (symbols).  Because of the six-fold symmetry, only the positive 
half of the first sextant is shown. (The negative half is mirror-symmetric.)        
}
{\label{fig:lt111}}
\end{figure}
%%%%%%%%%%%%%%%%%%%%%%%%%%%%%%%%%%%%%%%%%%%%%%%%%%%%%%%%%%%%%%%%%%%%%%%%%%%%%%%%%

\indent Just as for the stiffness, we systematically determine $X$ and its first two derivatives by 
differentiating  
the exact solution, Eqs.~(\ref{eq:ziaSol}-\ref{eq:psiZero}),
\begin{eqnarray}
  \label{eq:xlt}
  X &\equiv& \frac{a_{||}\beta(0)}{k_B T} = 2 \cosh^{-1}\left(\frac{y-1}{2}\right),\\
   X'&=&0, \\
   X'' &=& \frac{2 y \sqrt{y^2-2y-3}}{3(y-1)} - X.   
\end{eqnarray}
The last equation can be rearranged to find the reduced stiffness at $\theta=0$, 
as expressed earlier in Eq.~(\ref{eq:xtri}).  With these parameters in hand, we 
compare our approximation for 
the full orientation dependence of the reduced line tension with the exact, 
numerically evaluated solution in Fig.~\ref{fig:lt111}.  For the critical angle, 
we use Eq.~(\ref{eq:critang111}).  As before, the fit works remarkably well at 
temperatures as high as $T_c/5$.  

\section{\{001\} Surfaces with NN and NNN Interactions}

For \{001\} surfaces with just NN interactions, an exact, explicit form for the 
full orientation dependence of the line tension was first determined by Abraham and Reed \cite{abraham}.  
For such  
surfaces, however, NNN interactions are often significant \cite{tjs1}, so it is 
desirable to find a solution including their effects.  We denote by $R$ the ratio of NNN to NN adatom interaction strengths; the latter is assumed to be attractive (negative), so a positive $R$ 
indicates that the NNN interaction also is.

Although no exact solution to the Ising model with both NN and NNN interactions exists, the solid-on-solid (SOS) model provides an 
excellent approximation at reasonable temperatures ($\sim T_c/2$ based on our comparisons with the imaginary path weight random-walk method 
developed by the Akutsus \cite{akutsu}).  This model can be solved exactly \cite{tjs1}, yielding 
the following implicit form for the reduced line-tension:  
\begin{equation}
  \label{eq:ltexact001}
  \frac{\beta(\theta) a_{||}}{k_B T} = \left[ \rho(\theta) \sin \theta + g(\rho(\theta)) \right] \cos \theta,
\end{equation}
where $\rho(\theta)$ is found by inverting
\begin{equation}
  \label{eq:rho001exact}
\tan \theta   = \frac{ 2 \sinh \rho \, \sinh S}{(\cosh S - \cosh \rho )
\left[ 2 \sinh S - (\cosh S - \cosh \rho)(y+1)\right]},
\end{equation}
while $g(\rho)$ is 
\begin{equation}
  \label{eq:g}
  g(\rho)=S-\ln \left( \frac{y+1}{y-1} + \frac{2}{1-y}\frac{\sinh S}{\cosh S - \cosh \rho} \right).
\end{equation}
Here $y \equiv 1- 2 z^R$, $S\equiv -(R+1/2)\ln z$, 
$z\equiv(1+\sqrt{2})^{-2T_c/T} = \exp(-2\epsilon_k/k_BT)$, while $T_c$ is the critical temperature for $R=0$ (just NN interactions):
 
\begin{equation}
  \label{eq:kinke100}
\frac{\epsilon_k}{k_B T_c} = \ln (1+\sqrt{2}),
\end{equation}
\noindent  where the kink energy $\epsilon_k$ now refers to a close-packed step on an \{001\} surface.
We will utilize the exact, implicit solution Eqs.~(\ref{eq:ltexact001}-\ref{eq:g}) to determine the parameters required to find 
an explicit approximation for the stiffness and line tension below.

\subsection{Step Stiffness}

To begin, we let $X(\theta) \equiv k_B T/(\tilde{\beta}(\theta) a_{||})$.  
The symmetry of 
\{001\} surfaces  
require $X(\theta)$ be four-fold symmetric.  Accounting for just geometrically 
forced kinks, the reduced 
inverse stiffness is well 
approximated in the first quadrant ($-\pi/4$ to $\pi/4$) 
for $|\theta| > \theta_c$ by the following function \cite{tjs1}:
\begin{equation}
  \label{eq:fstiff100}
  f(\theta) = \frac{\sin(2 \theta)}{2}\sqrt{1- y \sin(2 \theta)}.
\end{equation}
By differentiating Eq.~(\ref{eq:fstiff100}), $f'$ and $f''$ are easily obtained.  

To determine $X$, $X'$, and $X''$ (and, potentially, any higher order derivatives), we utilize  
the exact solution of the NNN SOS model.  
Eq.~(\ref{eq:rho001exact}), for example, implies that $\rho_0 =0$ when $\theta = 0$.   
With some effort, it can be shown that 
\begin{eqnarray}
  \label{eq:dx100}
\! \!  X &\! =\! & \frac{ 2 \sinh S}{(\cosh S \! -\!  1)\left[2 \sinh S \! -\! (\cosh S \! -\! 1) (y\! +\! 1)\right]  }\\
\! \!    X' &\! =\! & 0 \\
\! \!    X'' &\! =\! & \frac{1}{X}\frac{2 \cosh S \! + \! 1}{\cosh S \! -\! 1} 
-4 \left[ \frac{\cosh S \! -\!  1}{\sinh S}\frac{y\! +\! 1}{2}+X \right]\! .
\end{eqnarray}
As required by symmetry, $X'=0$.  

%%%%%%%%%%%%%%%%%%%%%%%%%%%%%%%%%%%%%%%%%%%%%%%%%%%%%%%%%%%%%%%%%%%%%%%%%%%%%%%%%
\begin{figure*}[t]
\begin{minipage}[t]{8cm}
\begin{center}
\includegraphics[width=8cm]{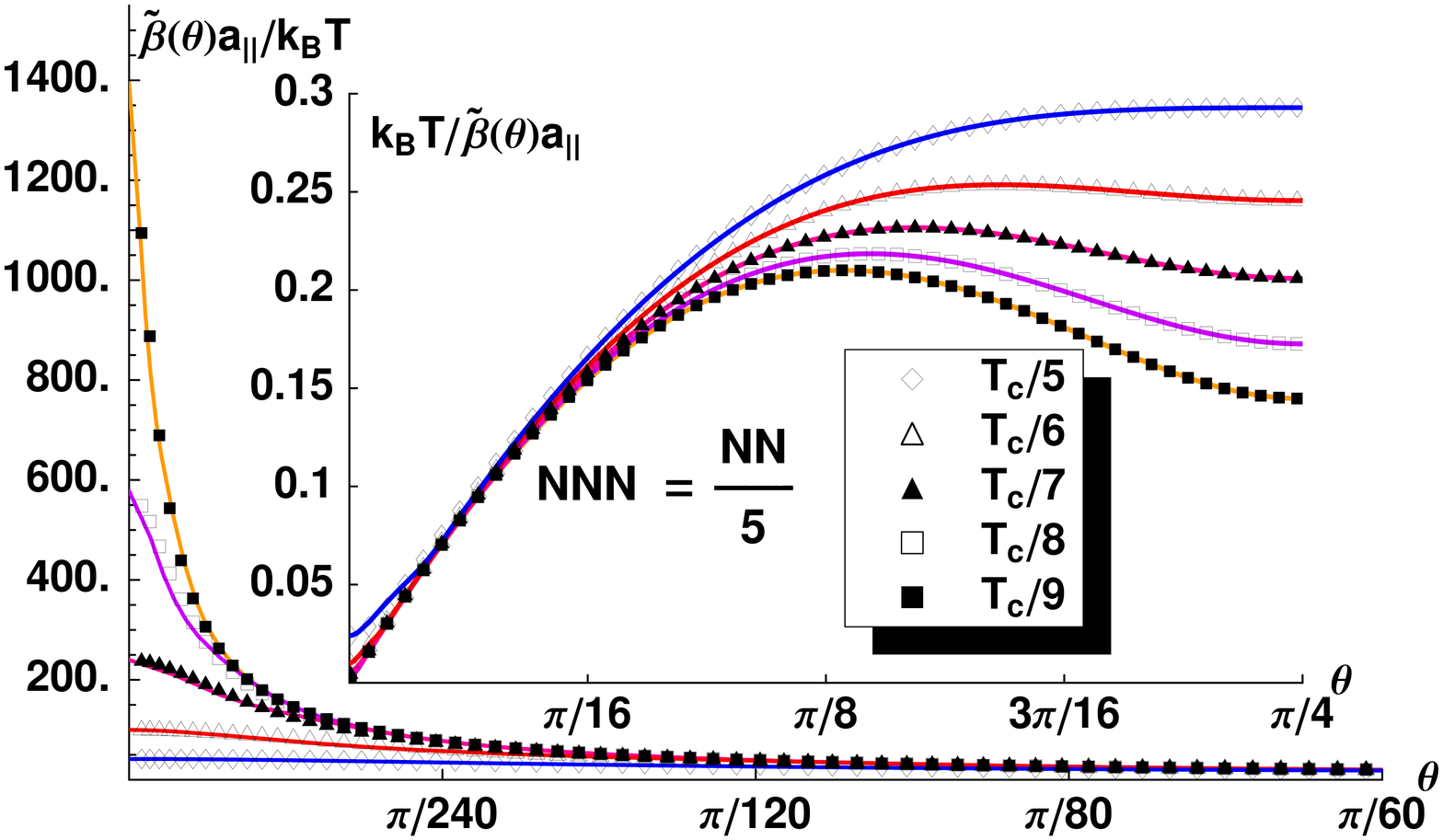}\\
\includegraphics[width=8cm]{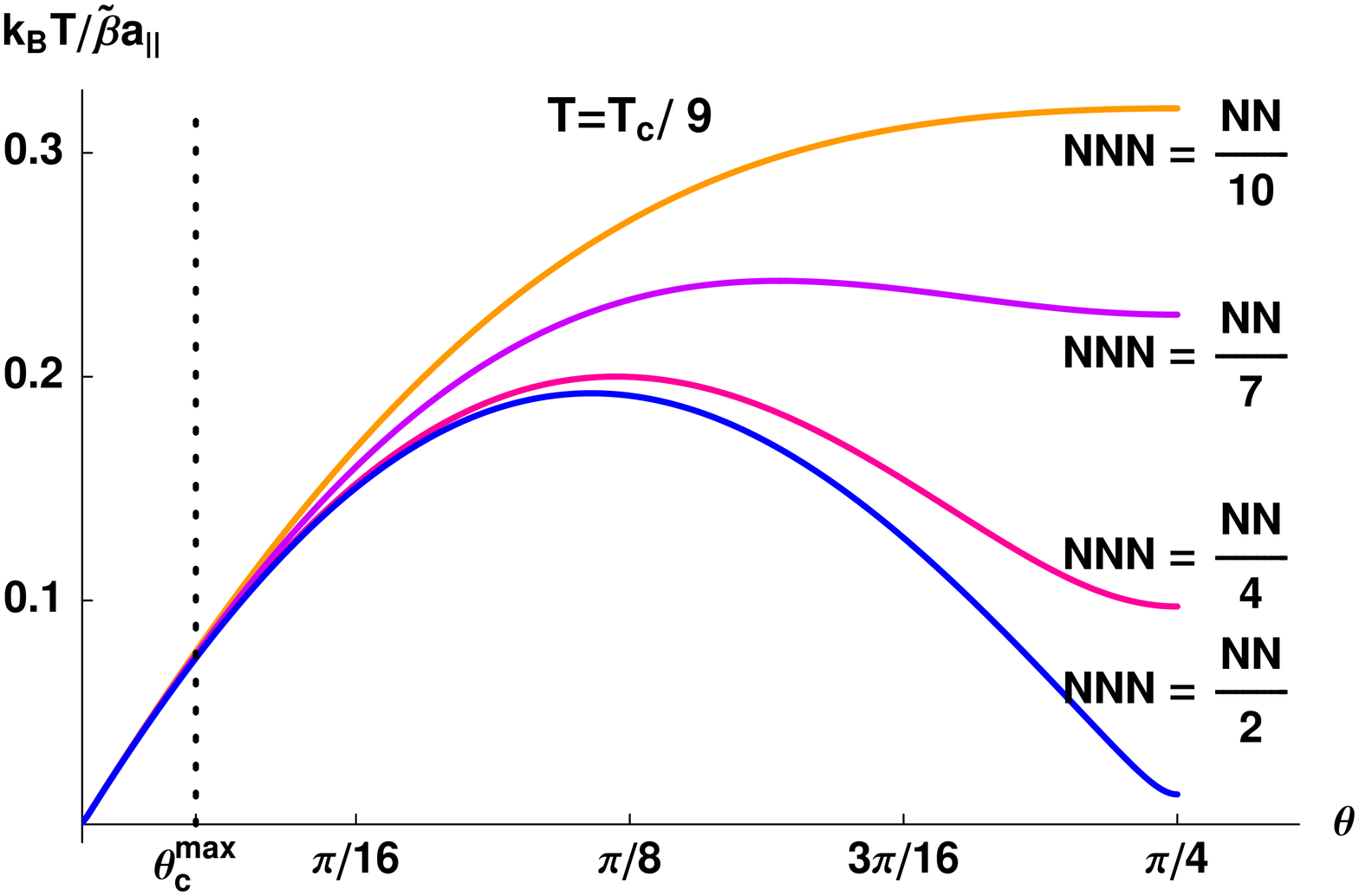}
\end{center}
\end{minipage}
%\hfill
\hspace{1cm}
\begin{minipage}[t]{8cm}
\begin{center}
\includegraphics[width=8cm]{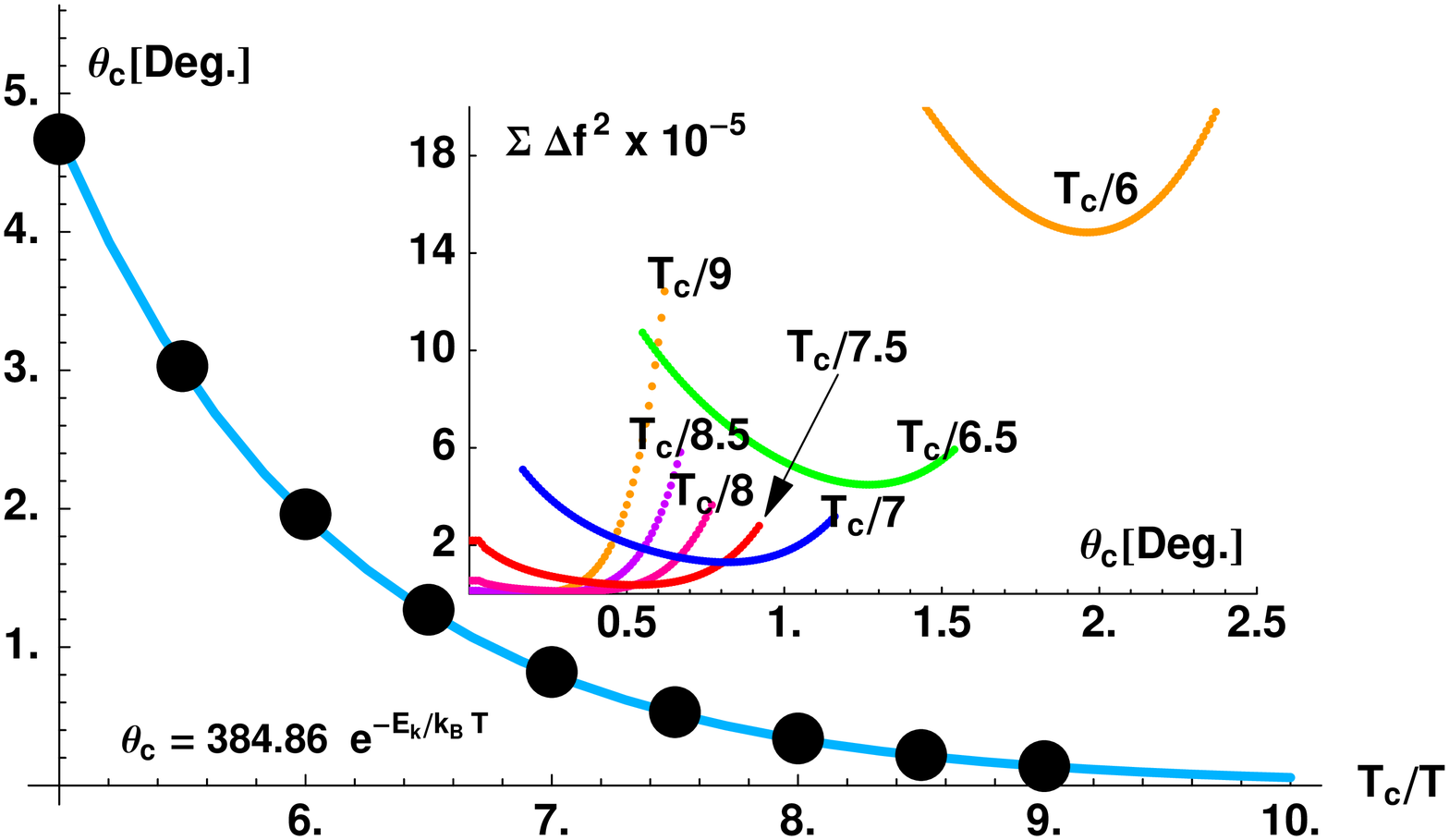}\\
\includegraphics[width=8cm]{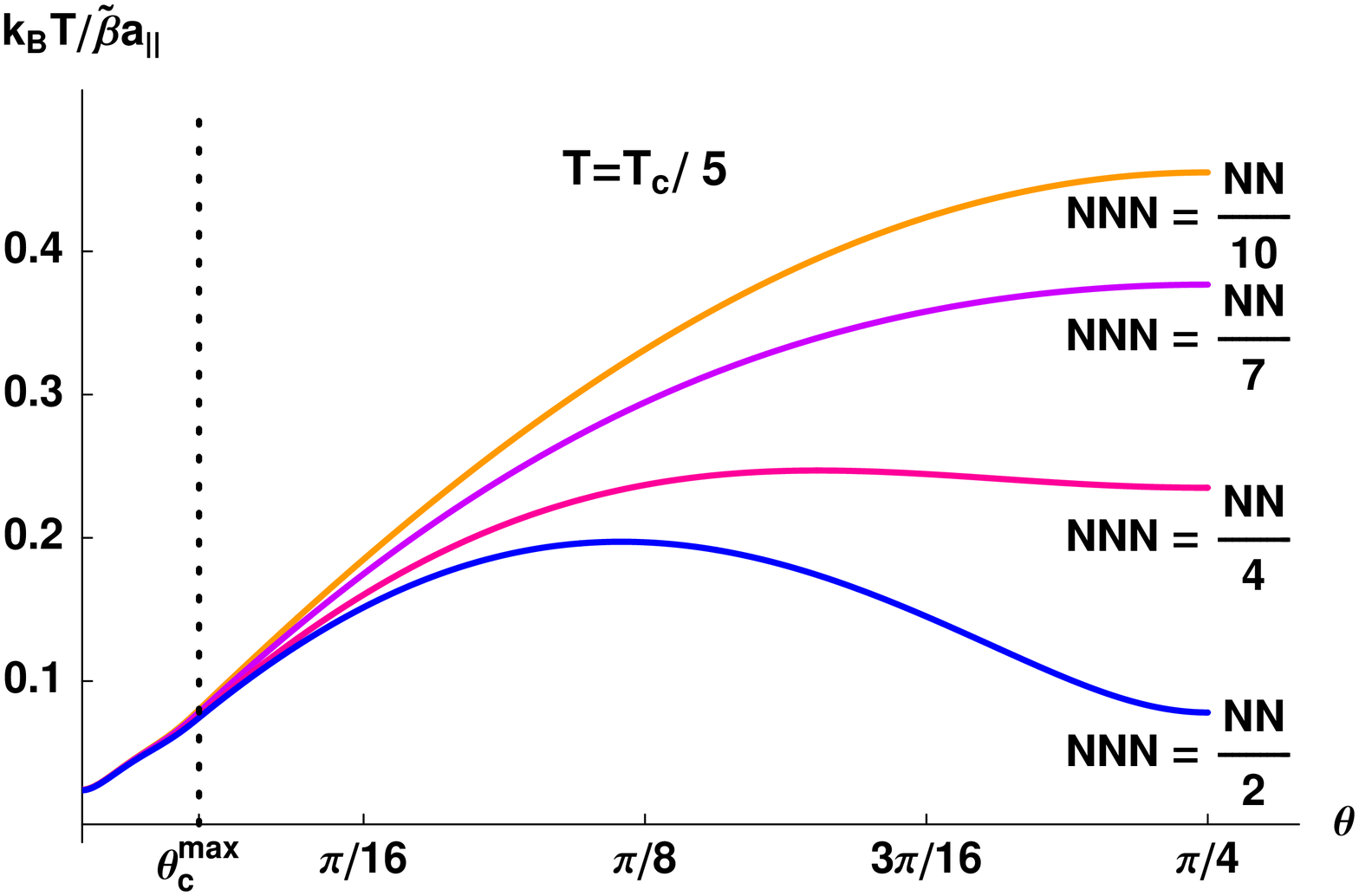}
\end{center}
\end{minipage}
\caption{(Color online) In the upper-left plot the 
orientation dependence of the explicit approximation  
for the \{001\} step stiffness (solid lines) and its inverse (inset, solid lines) are  
compared to the exact, implicit solutions (symbols).  Because of the four-fold 
symmetry of the solution, only the positive half of the first quadrant is shown 
(the negative half is mirror-symmetric).
The upper-right plot shows the values used for $\theta_c$ (solid dots) in the 
construction of the upper-left figure and the  
corresponding exponential decay fit (solid line) good over the temperature range of interest.
The fit is expressed in terms of the kink energy $\epsilon_k$ which is related   
to $T_c$ by Eq.~(\ref{eq:kinke100}).
The inset shows the sum of squared vertical deviations ($\sum \hat{\chi}^2$) 
versus angle in a least square fit 
for $\theta_c$.
At each temperature, $\theta_c$ is the angle that 
minimizes this sum.  
The two lower plots show 
the \{001\} inverse stiffness for a variety of different 
$R$ at two temperatures, $T_c/9$ and $T_c/5$ (the extremum of the 
temperature range of interest).  Notice that for a given temperature, all 
curves align at an angle greater than the largest critical angle $\theta_c^{max}$.
This behavior means $\theta_c$, practically speaking, 
does not depend on $R$ at these temperatures.}
{\label{fig:allAngSti100}}
\end{figure*}
%%%%%%%%%%%%%%%%%%%%%%%%%%%%%%%%%%%%%%%%%%%%%%%%%%%%%%%%%%%%%%%%%%%%%%%%%%%%%%%%%

Combining the functional forms for $f$, $X$, and their derivatives with 
Eqs.~(\ref{eq:expand1}-\ref{eq:bcfinal}), we can plot the inverse step stiffness and compare it 
to the numerically evaluated exact solution, just as before.  
We show this comparison in Fig.~\ref{fig:allAngSti100}, where 
$\theta_c$ was determined by doing least square fits to the numerically evaluated 
exact solution (with $R = 1/5$).  
The agreement shown in Fig.~\ref{fig:allAngSti100} is excellent at low-temperatures and 
is very reasonable at temperatures all the way 
up to $T_c/5$, as was the case for the \{111\} solution.

Although it was not initially obvious, 
the relative size of the NNN interaction $R$ has little effect on $\theta_c$.  This fortuitously implies  
that a single $\theta_c$ works for all values of $R$, as depicted in the lower plots of 
Fig.~\ref{fig:allAngSti100}.  

With this in mind, the values used for $\theta_c$ were determined just as they were 
for the \{111\} case, but with $R=1/5$.  These are shown in the 
upper-right plot of Fig.~\ref{fig:allAngSti100}, 
as well as a simple fit that is accurate over 
the temperature range of interest:  
\begin{equation}
  \label{eq:critang100}
  \theta_c(T) \approx 6.72\, e^{-\epsilon_k/k_B T} = 385[{}^\circ] (1+\sqrt{2})^{-T_c/T}.
\end{equation}

\noindent As for Eq.~(\ref{eq:critang111}), the second form uses the units in Fig.~\ref{fig:allAngSti100}, reexpressing the prefactor in degrees and the exponent in $T_c/T$.  Again, the Arrhenius decay is anticipated since $\theta_c$ 
represents the angle below which thermally activated kinks on close-packed segments become important. 

Finally, we point out  
that the \{001\} step stiffness is much more anisotropic than its \{111\} counterpart.  In fact, at 
$T_c/6$ the anisotropy is as large as the \{111\} anisotropy at $T_c/9$.  Furthermore, 
$\theta_c$ is less sensitive to temperature than its \{111\} counterpart.  This follows from 
the relative ease of thermally activating kinks on \{111\} steps, requiring only the 
breaking of one NN bond, as compared to two for \{001\} steps.  For \{111\} steps, then, the 
angle $\theta_c$ below which thermally activated kinks become important is larger than for \{001\} steps.        

However, there is no need to include an ``off-angle" correction $\Delta$ as was needed for the \{111\} case, at least in the case of just NN interactions ($R$=0, $y\! =\! -1$).  In that case one can readily find the difference between the stiffness of the exact result \cite{rottman,akutsu} and $f(\pi/4)$ from Eq.~(\ref{eq:fstiff100}):  

\begin{eqnarray}
  \label{eq:Delt100}
  \Delta_{100} &=&
  \frac{1}{\sqrt{2}}
  \left[\frac{1}{\sqrt{1\! -\! 4\, {\rm sech}^2(\epsilon_k/k_B T)
 \tanh^2(\epsilon_k/k_B T)}}
-1\right] \nonumber \\
  &=&\frac{2\sqrt{2}\sinh^2(\epsilon_k/k_B T)}{\cosh(2\epsilon_k/k_BT)-3}. 
\end{eqnarray}
\noindent Over the range of temperatures of interest here, numerical evaluation shows  $\Delta_{100}$ is negligible.

\subsection{Step Line Tension}
We proceed as usual, letting $X(\theta)\equiv \beta(\theta) a_{||}/k_B T$.  
The contribution from geometrically forced kinks is found by solving the low-temperature 
form of Eq.~(\ref{eq:rho001exact}), which becomes quadratic in $e^{\rho-S}$.  Solving gives  
\begin{equation}
  \label{eq:ltrho001}
  e^{\rho-S}=\frac{\sqrt{1-y \sin(2 \theta)}+y \sin \theta - \cos \theta}{(1+y) \sin \theta}.
\end{equation}
Plugging this into Eq.~(\ref{eq:ltexact001}) yields an 
excellent approximation $f(\theta)$ for the reduced line tension $X(\theta)$ valid in the first 
quadrant ($-\pi/4$ to $\pi/4)$ for $|\theta|>\theta_c$:  
\begin{eqnarray}
  \label{eq:ltlt001}
  &&\hspace{-4mm}f(\theta) = 
    \cos \theta \left[S + \ln  \frac{(1\! -y) \left(\sin \theta \!  +\!  \cos \theta\!  -\!  
                     \sqrt{1\! -y\!  \sin (2 \theta)} \right)} 
   {(1\! +\! y)\left( \sin \theta\!  -\!  \cos \theta \! +\!  \sqrt{1\! -\! y \sin (2 \theta)} \right) }  \right]  \nonumber \\
  &&  + \sin \theta \left[ S+\ln \frac{\sqrt{1-y \sin(2 \theta)}+y \sin \theta - \cos \theta}
                                     {(1+y) \sin \theta}   \right] 
\end{eqnarray}
Differentiating twice straightforwardly gives $f'$ and $f''$.  Eq.~(\ref{eq:ltlt001}) can be written more compactly by defining and inserting $w(\theta,y) \equiv \left[\cos \theta -   \sqrt{1 - y \sin (2 \theta)}\right]/\sin \theta$, as done in Table~5.1.

This leaves $X$ and its derivatives.  They too can be explicitly determined 
from the exact solution.  
Setting both $\theta=0$ and $\rho=0$ (as Eq.~(\ref{eq:rho001exact}) demands) 
in Eq.~(\ref{eq:ltexact001}), we find $X$:
\begin{eqnarray}
  \label{eq:x001}
  X&=&g(0) \nonumber \\
   &=& S - \ln \left( \frac{y+1}{y-1} + \frac{2}{1-y}\frac{\sinh S}{\cosh S - 1}  \right).
\end{eqnarray}
Similarly, it can be shown that 
\begin{equation}
  \label{eq:dx100lt}
   X'= 0,  
\end{equation}
\begin{equation}
  X''\!  =\!  \frac{(\cosh S \!  -\!  1)[2 \sinh S \! -\!  (\cosh S \!  -\!  1)(y\! +\! 1)]}{2 \sinh S}\!  -\! X. 
\end{equation}
This last equation can be rearranged to give the reduced step stiffness, as previously 
written in Eq.~(\ref{eq:dx100}).  

\begin{figure}[t]
\begin{center}
\includegraphics[width=8.2cm]{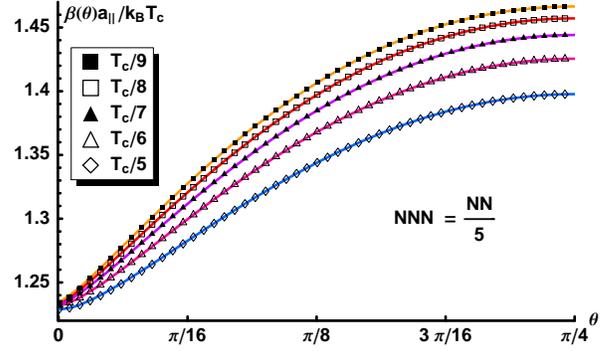}
\end{center}
\caption{(Color online) The orientation dependence of the explicit approximation 
for the  \{001\} line tension (solid lines) is compared with the numerically evaluated 
exact result (symbols).  Because of the four-fold symmetry, only the positive 
half of the first quadrant is shown (the negative half is mirror-symmetric).        
}
{\label{fig:lt100}}
\end{figure}

By combining the functional forms for $f$ and $X$ and their derivatives with 
Eqs.~(\ref{eq:expand1}-\ref{eq:bcfinal}), we can plot the reduced line tension 
and compare it 
to the numerically evaluated exact solution.  
We show this comparison in Fig.~\ref{fig:lt100}, where $\theta_c$ was determined 
from Eq.~(\ref{eq:critang100}) and $R=1/5$ (other values yield equally good agreement).
As before, the approximation works well at temperatures up to $T_c/5$ (and, in this 
case, perhaps even higher).     

%%%%%%%%%%%%%%%%%%%%%%%%%%%%%%%%%%%%%%%%%%%%%%%%%%%%%%%%%%%%%%%%%%%%%%%%%%%%%%%%%
%%%%%%%%%%%%%%%%%%%%%%%%%%%%%%%%%%%%%%%%%%%%%%%%%%%%%%%%%%%%%%%%%%%%%%%%%%%%%%%%%

\section{Summary and Concluding Remarks}

\setlength{\extrarowheight}{6pt}
\begin{table*}[t]
\centering
\begin{tabular}{|>{\centering}m{0.2in}|| >{\centering}m{2.3in} m{2.05in}| }
\hline
\multicolumn{3}{|c|}{ \textbf{EXPLICIT APPROXIMATION FOR STIFFNESS \& LINE TENSION} }  \\[2ex]
\multicolumn{3}{|c|}{  
$X(\theta) := \left\{ \!
\begin{array}{cc}
           {\displaystyle \sum_{n=0}^{2 N-1}} a_n~ \theta^n, & \mbox{$\theta < \theta_c$} \nonumber \\
         f(\theta), & \mbox{$\theta \geq \theta_c$} 
\end{array} 
\right.
$\quad
$\begin{array}{ll}
a_0=X & a_3=\frac{20 (f-X)-8 f' ~\theta_c + (f''-3 X'') ~\theta_c^2}{2 ~\theta_c^3} \\
a_1=0 & a_4=\frac{-30 (f-X)+14 f' ~\theta_c - (2 f''-3 X'') ~\theta_c^2}{2 ~\theta_c^4} \\
a_2=\frac{X''}{2} & a_5= \frac{12 (f-X) -6 f' ~\theta_c + (f''- X'') ~\theta_c^2}{2 ~\theta_c^5}
\end{array}
$
} \\[7ex]
\hline \hline 
&\multicolumn{2}{c|}{\color{DarkRed} \textbf{\{111\} Surfaces with NN Interactions} } \\
\color{DarkRed}$\mathbf{\downarrow}$ &\multicolumn{2}{c|}
{\color{DarkRed}$\theta_c[^\circ]=642~e^{-\beta_{\rm B}{\textstyle\epsilon_k}}$, \quad
$z=3^{-T_c/T}=e^{-2\beta_{\rm B} {\textstyle\epsilon_k}}$, \quad $y=\sqrt{(3z+1)/z(1-z)}$}
 \\[2ex]
\cline{2-3}
&\multicolumn{1}{c}{\textbf{Stiffness ($X(\theta) \approx k_{\rm B} T/ a_{||} \tilde{\beta}$)}}
&\multicolumn{1}{|c|}{\textbf{Line Tension ($X(\theta)\approx a_{||} \beta/k_{\rm B} T$)}} 
\\ [1ex] 
\hline 
$\color{DarkRed}X$&\color{DarkRed}$ \frac{3(y-1)}{2 y \sqrt{y^2-2y-3}}$
& \multicolumn{1}{|c|}{\color{DarkRed}$ 2 \cosh^{-1}(\frac{y-1}{2})$}\\

$\color{DarkRed}X''$&\color{DarkRed}$ \frac{y^3-2y^2-15y+36}{2(y-1)\sqrt{y^2-2y-3}}$
& \multicolumn{1}{|c|}{\color{DarkRed}$\frac{2 y \sqrt{y^2-2y-3}}{3(y-1)} - X$}\\

$\color{DarkRed}f(\theta)$
&\color{DarkRed}$ \frac{1}{2\sqrt{3}} \left( \sin (3 \theta) + 
        \frac{3+y^2}{\sqrt{y^4-10 y^2+9}}-1\right)$
& \multicolumn{1}{|c|}{\color{DarkRed} $ -\eta_{+} \ln (z \eta_+) + 
                  \eta_- \ln \eta_- + \eta_0 \ln \eta_0 \; ^*$} 
\\ \hline  \hline

 \color{blue}$X$&  \color{blue}$ \frac{ 2 \sinh S}{(\cosh S - 1) \left[ 2 \sinh S - (\cosh S - 1) (y+1)\right]  } $  
& \multicolumn{1}{|c|}{ \color{blue}$ S - \ln \left( \frac{y+1}{y-1} + \frac{2}{1-y} \frac{\sinh S}{\cosh S-1}\right)  $}\\

$ \color{blue}X''$& \color{blue}$ \frac{1}{X} \frac{2 \cosh S +1}{\cosh S - 1} 
-4\left[ \frac{\cosh S-1}{\sinh S}\frac{y+1}{2}+X \right]$
& \multicolumn{1}{|c|}{ \color{blue}$ \frac{(\cosh S - 1)[2 \sinh S - (\cosh S - 1)(y+1)]}{2 \sinh S}-X$}\\

$ \color{blue}f(\theta)$
& \color{blue}$ \frac{\sin(2 \theta)}{2}\sqrt{1- y \sin(2 \theta)} $
& \multicolumn{1}{|l|}{ \color{blue}$ 
      \qquad \cos \theta \left[S  \! + \!  \ln  \frac{(1-y) \left(1 + w(\theta,y) \right)} 
   {(1+y)\left( 1 - w(\theta,y) \right) }  \right] +$ }\\
 &  & \multicolumn{1}{|c|}{ \color{blue}   $\qquad +   \sin \theta \left[ S \! + \! \ln \frac{y - w(\theta,y)} {(1+y) }   \right] \;^\dagger $}
  \\ \hline
 $ \color{blue}\mathbf{\uparrow}$   &\multicolumn{2}{c|}{ \color{blue} \textbf{\{001\} Surfaces with NN and NNN (= R$\times NN$) Interactions} } \\
&\multicolumn{2}{c|}{ \color{blue}$\theta_c[^\circ]\! =\! 385~e^{-\beta_{\rm B}{\textstyle\epsilon_k}}\! , \; 
   z\! =\! e^{-2 \beta_{\rm B}{\textstyle\epsilon_k}}\! =\! (1\! +\! \sqrt{2})^{-\frac{2 T_c}{T}} \! , \; 
%    z\! =\! e^{-2 \beta_{\rm B}{\textstyle\epsilon_k}}\! =\! (1\! +\! \sqrt{2})^{-2 T_c/T}, \; 
   S\! =\! (1\! +\! 2R) \beta_{\rm B}{\textstyle\epsilon_k}, \;  y\! =\! 1\! -\! 2z^R$       }
   \\[2ex]
\hline 
\multicolumn{3}{c}{$\color{DarkRed} {}^*\eta_{\pm} \equiv \cos \theta \pm \frac{1}{\sqrt{3}} \sin \theta, \; 
\eta_0\equiv \frac{2}{\sqrt{3}} \sin \theta   \hspace{30pt} \color{blue} ^\dagger w(\theta,y)  \equiv  \cot \theta -  \csc \theta \sqrt{1\! -\! y \sin (2 \theta)}$} 
\end{tabular}
\vspace{5mm}
\caption{(Color online) Summary of results for approximants of dimensionless inverse stiffness and line tension.  $X \equiv X(0)$, while $f\equiv f(\theta_c)$; $\beta_{\rm B}\! \equiv\! (k_{\rm B}T)^{-1}$, the subscript needed to distinguish it from the line tension.  The upper part of the table (dark red) refers to the steps on the hexagonal-lattice face, with just NN interactions.  The lower part (blue) refers to the square-lattice face; by setting $R$=0, one retrieves the simpler formulas for just NN interactions.}
\label{tab:summary}
\end{table*}

We have constructed explicit, twice-differentiable approximants for the full anisotropy  
of step stiffness and line tension on both 
\{001\} and \{111\} surfaces of fcc crystals, the metallic systems that have been subjected to the greatest scrutiny with regard to island properties \cite{bonz,MK}.  Our expressions are accurate over a broad range of 
experimentally relevant temperatures; they fail only when the stiffness is nearly isotropic, i.e., 
when their use is no longer required.  Implementation
into continuum simulations is straightforward and efficient.  They are much more usable than numerically extracting solutions from the underlying 6th-order equations, and more flexible and convenient than constructing immense look-up tables as functions of angle and temperature from such a procedure.
Our expressions are greatly superior to conventional explicit formulas for step stiffness and 
line tension,  
which usually take the form of simple sinusoidal variation that neither carry temperature dependence nor 
accurately capture the 
anisotropy (extreme for the step stiffness) observed at lower temperatures.  
For clarity and convenience, we summarize our results in Table \ref{tab:summary}.  

We have implemented these formulas into state-of-the-art finite-element simulations, where we have tested them by numerically determining the equilibrium shape of two-dimensional islands at various temperatures \cite{TJStaoFH}.  
Explicitly, this is done by 
finding the shape that minimizes the chemical potential (proportional to the product of the step curvature and stiffness) of a single, island-bounding step.  We have also used these simulations to model island relaxation from an arbitrary shape to the 
equilibrium shape and are currently using them to model recent experiments monitoring the relaxation of 
depinned Ag(111) steps \cite{TJStaoFH}.

\section*{Acknowledgements}
We thank D.\ Margetis and A.\ Voigt for helpful discussions.


\end{document}